\documentclass[letterpaper, 10 pt, conference]{ieeeconf}

\def\BibTeX{{\rm B\kern-.05em{\sc i\kern-.025em b}\kern-.08em
    T\kern-.1667em\lower.7ex\hbox{E}\kern-.125emX}}

\IEEEoverridecommandlockouts
\usepackage{graphics}
\usepackage{epsfig} 
\usepackage{mathptmx} 
\usepackage{times} 
\usepackage{amsmath} 
\usepackage{amssymb} 
\usepackage{xcolor}
\usepackage{bm}
\newtheorem{proposition}{Proposition}
\newtheorem{assumption}{Assumption}
\newtheorem{definition}{Definition}
\newtheorem{problem}{Problem}

\title{\LARGE \bf  From Inference to Control: Structure-Guided Control of \\ Hypergraph Dynamics}

\author{Robin Delabays$^{1}$, Can Chen$^{2}$, Florian Dörfler$^{3}$, Yuanzhao Zhang$^{4}$, and Giulia De Pasquale$^{5}$
\thanks{*RD was supported by the Swiss National Science Foundation under grand nr. 200021\_215336}.%
\thanks{$^{1}$Robin Delabays is with the University of Applied Sciences of Western Switzerland (HES-SO), Sion, Switzerland. {\tt\footnotesize robin.delabays@hevs.ch}}
\thanks{$^{2}$Can Chen is with the School of Data and Information Sciences, the Department of Mathematics, and the Department of Biostatistics, University of North Carolina at Chapel Hill, Chapel Hill, USA. {\tt\footnotesize canc@unc.edu}}
\thanks{$^3$Florian Dörfler is with the Department of Information Technology and Electrical
Engineering at ETH Zürich, Zürich, Switzerland. {\tt\footnotesize dorfler@ethz.ch}}
\thanks{$^{4}$Yuanzhao Zhang is with the Department of Physics and Astronomy at the University of Rochester, Rochester, New York, USA.{\tt\footnotesize yuanzhao.zhang@rochester.edu}}
\thanks{$^5$Giulia De~Pasquale is with the Control Systems group in the Electrical Engineering Department and the Eindhoven AI Systems Institute at Eindhoven Univeristy of Technology, Eindhoven, The Netherlands.{\tt\footnotesize g.de.pasquale@tue.nl}}}

\begin{document}

\maketitle
\thispagestyle{empty}
\pagestyle{empty}

\begin{abstract}
Controllability determines whether a system’s state can be guided toward any desired configuration, making it a fundamental prerequisite for designing effective control strategies. In the context of networked systems, controllability is a well-established concept. However, many real-world systems, from biological collectives to engineered infrastructures, exhibit higher-order interactions that cannot be captured by simple graphs. Moreover, the interaction structures might be unknown and difficult to measure directly.
Here, we close this gap by combining hypergraph inference with the identification of controllable nodes.
Building on the inferred structure, we design a controller that, given a set of controllable nodes, steers the system toward a desired configuration. We formulate analytical controllability guarantees for polynomial systems. For non-polynomial dynamics on hypergraphs, we propose a heuristic method for identifying controllable nodes and validate the proposed approach using Kuramoto oscillators.
\end{abstract}

\section{Introduction}

\emph{Motivation.} 
Networked systems provide a powerful framework for describing the collective dynamics of interacting agents, from power grids \cite{power_system_general}  to biological  \cite{biological_sys_general} and social systems \cite{social_net}. 
However, in many real systems, interactions that naturally occur among groups of agents cannot be reduced to a set of pairwise interactions, for example, in neuronal assemblies \cite{hyper_brain}, ecological communities \cite{hyper_ecology}, and social groups \cite{alvarez2021evolutionary} where the state of an individual depends simultaneously on multiple others. These higher-order interactions are more accurately captured by hypergraphs or simplicial complexes, which generalize networks by allowing multi-body connections~\cite{bianconi2021higer}. 
The ability to understand and control such systems is crucial for ensuring stability and robustness. Yet, in many practical scenarios, the underlying interaction structure is not directly observable and must be inferred from limited data. This lack of full structural knowledge challenges traditional control approaches, which typically assume a known network topology. 

By integrating structure inference and control design within a unified framework, we aim to develop principled tools for identifying and steering higher-order dynamical systems, enabling effective interventions even when the governing interactions are only partially known.

More concretely, extensions of the well studied Kuramoto model to include three-body or higher-order interactions have revealed that these couplings can make the synchronized state’s basin of attraction smaller yet deeper, raising the critical coupling threshold while enhancing robustness once synchronization is achieved \cite{zhang2023higher}. These findings highlight that higher-order structures fundamentally reshape the landscape of collective dynamics. 

The coupling strength plays a central role in determining the collective behavior of oscillator networks \cite{pnas_sync_dorfler}. It governs the transition from incoherence to synchrony and critically shapes the stability, robustness, and transient dynamics of the ensemble.  However, in most practical scenarios, coupling strengths are not directly measurable, as the interactions between oscillators are often mediated by latent physical, chemical, or informational processes. Instead, only phase measurements are typically accessible. 

Similarly, the generalized Lotka-Volterra model~\cite{goel1971volterra} is typically used to describe the evolution of populations abundances in an ecosystem. 
While in macroscopic ecosystems the type of interaction between two species (prey-predator, collaboration, competition) may be well documented, the actual strength of this interaction is usually much more uncertain. 
In microscopic ecosystems (involving mostly bacteria), the vast majority of inter-species interaction is mostly unknown. 
In such systems, the literature relies mainly on random matrix theory~\cite{ros2023generalized}.

These limitations motivated us to design a closed loop algorithm between our proposed Taylor-based Hypergraph Inference using SINDy (THIS) algorithm \cite{this}, able to estimate effective coupling parameters and interaction structures from time series data together with a control law in order to drive the system towards a desired state. Such approach, allows to identify hidden connectivity patterns, assess synchronization or steady state regimes, and design control strategies in real-world networks where the underlying coupling mechanisms remain partially unknown.

\emph{Literature Review.} 
The notion of structural controllability for linear time invariant systems evolving over network traces back to the 70s \cite{structural_controllability} with the work of Lin on "structural controllability". 
Both algebraic and graph theoretic conditions were provided to ensure that a system can be driven towards any point of the state space based solely on topological considerations. Further analytical tools to certify structural controllability of a network have been provided in \cite{barabasi_controllability}, where the authors identify the set of driver nodes with time-dependent control that can drive the system dynamics. 
The extension of the notion of controllability to hypergraphs is much more recent with Refs.~\cite{chen_controllability_hyper,ma2024controllabilitya,dong2024controllability}, which relies on tensor algebra and polynomial control theory~\cite{jurdjevic1985polynomial}. They provide theoretical results about the minimum number of control nodes to be controlled given specific hypergraph structures. 

\emph{Contributions.}We leverage our recently proposed identification algorithm, Taylor-based Hypergraph Inference using SINDy (THIS) \cite{this}, to infer coupling strengths in coupled oscillators. Building on the inferred model, we design a simple yet effective proportional control strategy that enables systems evolving over a hypergraph to reach a desired fixed point. We provide formal guarantees for a non-homogeneous hypergraph of polynomial systems to be structurally controllable. We demonstrate the proposed identification and control framework on a hypergraph of polynomial dynamics (generalized Lotka-Volterra model) as well as non-polynomial dynamics (higher-order Kuramoto model).

\emph{Article Organization.} The paper is organized as follows. Section \ref{sec:notation} introduces the notation and preliminary concepts used throughout the paper. Section \ref{sec:hyper_stab} presents the problem formulation for the design of a  state-feedback controller. Section \ref{se:algo} details the proposed algorithm design, while Section \ref{sec:numerics} demonstrates its effectiveness on a hypergraph of Kuramoto oscillators. Finally, Section \ref{sec:conclusions} concludes the manuscript.

\section{Notation and Preliminaries}\label{sec:notation}
We provide the notation and preliminary concepts that will be used throughout the manuscript.

\subsection{Hypergraphs}
For a matrix $A\in \mathbb{R}^{n\times n}$, let  $a_{i, j}$ be its $(i, j)$-th entry. 
A \emph{$p$-th order hypergraph} $\mathcal{H}$ is a tuple $\mathcal{H}(V,E,A^{(2)},...,A^{(p)})$, where $V=\{1,\dots,n\}$ indicates the set of nodes, $E^{(k)}\subset V^k$ is the set of $k$-th order (hyper)edges and $E = \bigcup_kE^{(k)}$ is the set of all (hyper)edges, and $A^{(k)}\in\mathbb{R}^{n\times \dots \times n}$ is the $k$-th order \emph{adjacency tensor}, such that a $k$-hyperedge $(i_1,i_2,\dots,i_k)$ is present in the hypergraph if and only if $A^{(k)}_{i_1,i_2,\dots,i_k}\neq 0$.

In this work, we consider \emph{directed} hypergraphs, meaning that each permutation of $k$ indices $(i_1,\dots,i_k)$ denotes a distinct directed hyperedge. 
The tuple $e=(i_1,i_2,\dots,i_k)$ indicates that nodes $i_2,\dots,i_k$ all influence node $i_1$'s dynamics.
We say that the node at the first index is the \emph{head} of a hyperedge and the other nodes are the \emph{tails}, e.g., in $e=(i_1,i_2,\dots,i_k)$, node $i_1$ is the head of $e$, and $i_2,\dots,i_k$ are the tails. 

\begin{figure}
    \centering
    \includegraphics[width=\columnwidth]{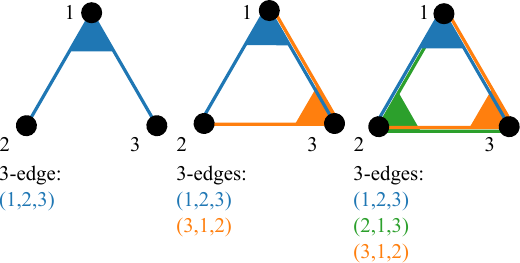}
    \caption{Examples of leaves in hypergraphs and the choice of control nodes. 
    In the left panel, nodes 2 and 3 are leaves and need to be controlled.
    In the middle panel, only node 2 is a leaf and needs to be controlled. 
    In the right panel, there is no leaf node and one can control any one of the three nodes. }
    \label{fig:loose-ends}
\end{figure}

We define a sequence of edges $(e_1,...,e_\ell)$ is a \emph{directed path} from node $i$ to node $j$ if $i$ is the head of $e_1$, $j$ is a tail of $e_\ell$, and the head of $e_k$ is one of the tails of $e_{k-1}$, for each $k\in\{2,...,\ell\}$. 
A node is said to be a \emph{leaf} if it is not the head of any edge. 
For instance, in the middle panel of Figure~\ref{fig:loose-ends}, node 2 is a leaf, while nodes 1 and 3 are not.
A hypergraph is said to be \emph{connected} if for any two nodes $i$ and $j$, there is either a path from $i$ to $j$ or a path from $j$ to $i$.

\section{Hypergraph Stabilization via Noise Rejection}\label{sec:hyper_stab}

Consider a hypernetworked system of arbitrary order subject to additive noise described as 
\begin{align}\label{eq:system}
    \dot{x}_i &= \xi_i + f(x_i) + \sum_j a_{ij}f^{(2)}(x_i,x_j) + \sum_{j,k}a_{ijk}f^{(3)}(x_i,x_j,x_k) + \cdots
\end{align}
where $f^{(k)}$ is the $k$-th-order interaction function and $\xi_i$ is the noise term at node $i$. 
We do not have any knowledge about the functions $f^{(k)}$, and we only have access to time-series trajectories $x(t)$ of \eqref{eq:system}.  We will assume that the autonomous system in~\eqref{eq:system} admits at least one stable equilibrium point and that it is operating in the proximity of one of them, as formalized by the following assumption.

\begin{assumption}
    The autonomous system in \eqref{eq:system} admits at least one locally stable equilibrium point $x^*$.
\end{assumption}

Note that, due to the noise term $\xi_i$, the system evolves randomly around the fixed point ${x}^*$. 
For a strong enough disturbance, the system may even escape the basin of attraction of ${x}^*$ and either reach another basin of attraction, or lose stability and diverge. 
Our goal in this manuscript is to design an algorithm that, given the state trajectories $x(t)$, identifies the hyperedges in the system and  a  droop control that dampens the excursions due to noise and prevents destabilization of the system. 
Our state feedback control problem is formally stated as follows:

\begin{figure*}
    \centering
\includegraphics[width=\textwidth]{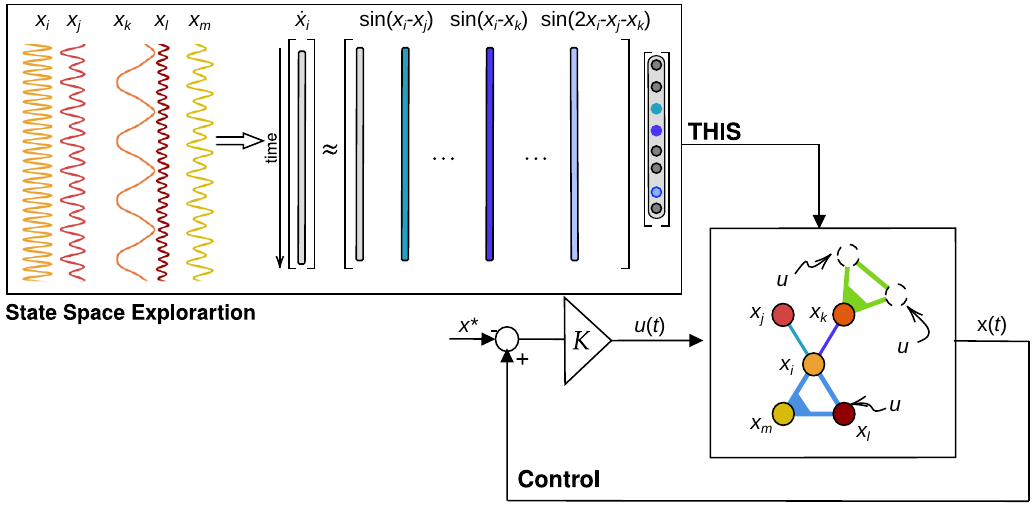}
    \caption{Algorithm process pipeline: In phase 1) trajectories are collected for a system operating in the vicinity of a locally stable equilibrium point for a time interval $[t_0,t_1]$; at time instant $t_2$ phase 2 starts, where THIS is applied to infer the hypergraph topology, after which, at phase 3) the minimal set of controllable nodes is detected and the system driven towards the desired equilibrium point $x^*$.}
    \label{fig:summary}
\end{figure*}

\begin{problem}\label{pb:1}
    Given state trajectories of the system \eqref{eq:system}, design a  state-feedback controller $u(x)$, so that \eqref{eq:system} converges to the closest equilibrium point $x^*$.
\end{problem}

This problem is of particular relevance in the context of chemical reactions, where the whole interaction structure may be complicated to grasp. 
Furthermore, the ability to drive the reaction towards a desired state, using minimal resources is valuable for many situations, such as optimizing catalytic processes in industrial chemistry to increase yield while reducing waste \cite{sheldon1997catalysis}.

\section{Identification of controllable nodes}\label{se:algo}
To address Problem~\ref{pb:1}, we first identify a set of nodes that can be used as control inputs. We consider separately the cases in which the dynamics are polynomial and non-polynomial. The identification procedure is based on the structure of the hypergraph and does not require explicit knowledge of the hyper-edge weights. Importantly, the procedure described below is a \emph{heuristic}: it provides a structurally motivated candidate set of controllable nodes, but it does not, in general, establish that the identified set is a minimum-cardinality control set or that the corresponding nonlinear system is controllable from every initial state.

\subsection{Polynomial systems}
We first consider the case in which the dynamics are polynomial.
A non-uniform hypergraph may contain interactions of different
cardinalities, resulting in polynomial terms of different degrees.
The corresponding control-affine polynomial system can be written as
\begin{equation}
    \dot{x}
    =
    F(x)+Bu
    =
    \sum_{k=2}^{d} F^{(k)}(x)+Bu
    =
    \sum_{k=2}^{d} A^{(k)}x^{k-1}+Bu,
    \label{eq:nonuniform-polynomial-system}
\end{equation}
where
    $F^{(k)}(x)=A^{(k)}x^{k-1},$
$A^{(k)}$ is the $k$th-order tensor associated with the
$k$th-order interactions, and $A^{(k)}x^{k-1}$ denotes its contraction
with $k-1$ copies of $x$. Thus,
$F^{(k)}:\mathbb{R}^{n}\rightarrow\mathbb{R}^{n}$ is a homogeneous
polynomial vector field of degree $k-1$. The matrix $B$ is the input
matrix, and $d$ denotes the highest interaction order. We assume that
$d$ is even. Note that $F^{(k)}$ denotes the aggregate polynomial vector
field associated with the $k$th-order interactions and is distinct from
the interaction function $f^{(k)}$ introduced in Section~III.

\begin{definition}[Structural Controllability]\label{def:struct_contr}
The polynomial system \eqref{eq:nonuniform-polynomial-system} is said to be structurally controllable if there
exists a realization $\{\bar{A}^{(k)}\}_{k=2}^{d}$ and $\bar{B}$,
consistent with the prescribed zero--nonzero patterns of
$\{A^{(k)}\}_{k=2}^{d}$ and $B$, for which the resulting polynomial
system is controllable.
\end{definition}

Let $H_d$ denote the subhypergraph associated with the highest-order
component $F^{(d)}$, together with the control inputs induced by $B$. 
The corresponding homogeneous subsystem is
\begin{equation}
    \dot{x}
    =
    F^{(d)}(x)+Bu
    =
    A^{(d)}x^{d-1}+Bu.
    \label{eq:highest-degree-subsystem}
\end{equation}
The  definition of structural controllability applies to a homogeneous polynomial system as a
special case in which only one homogeneous component is present. Notably, structural controllability of such homogeneous polynomial systems has
recently been studied in \cite{pickard2026structural}, where
graph-theoretic conditions are developed to characterize structural
controllability in terms of the underlying directed hypergraph and the placement of control inputs. In what follows, we extend these results
to the non-uniform polynomial system
\eqref{eq:nonuniform-polynomial-system} by relating its structural
controllability to that of its highest-order homogeneous subsystem
\eqref{eq:highest-degree-subsystem}.

\begin{proposition}
\label{prop:nonuniform-highest-degree}
If the homogeneous subsystem
\eqref{eq:highest-degree-subsystem} is structurally controllable, then
the complete non-homogeneous polynomial system
\eqref{eq:nonuniform-polynomial-system} is structurally controllable.
Consequently, any driver node set that renders $H_d$ structurally
controllable is sufficient for the complete non-uniform system.
\end{proposition}

\begin{proof}
By the structural controllability result for homogeneous polynomial
systems in~\cite{pickard2026structural}, structural controllability of
$H_d$ implies that there exists an admissible realization
$(\bar{F}^{(d)},\bar{B})$, consistent with the prescribed structure,
for which \eqref{eq:highest-degree-subsystem} is controllable.
Choose arbitrary nonzero lower-degree coefficients consistent with the
remaining structure of $H$ and define
\[
    \bar{F}(x)
    =
    \sum_{k=1}^{d}\bar{F}^{(k)}(x).
\]
For polynomial control systems with constant control vector fields and
odd highest degree, controllability is determined by the highest-degree
homogeneous component~\cite{jurdjevic1985polynomial}. In particular,
for constant vector fields $v_1,\ldots,v_d$, repeated Lie bracketing
gives, up to the Lie-bracket sign convention,
\[
    \operatorname{ad}_{v_d}\cdots
    \operatorname{ad}_{v_1}\bar{F}
    =
    D^d\bar{F}[v_1,\ldots,v_d]
    =
    D^d\bar{F}^{(d)}[v_1,\ldots,v_d],
\]
since $D^d\bar{F}^{(k)}=0$ for all $k<d$. Thus, the constant
directions generated by the highest-degree homogeneous component
remain available in the Lie algebra of the complete polynomial system.

Since $(\bar{F}^{(d)},\bar{B})$ is controllable, the corresponding
highest-degree controllability condition is satisfied. The polynomial
controllability result in~\cite{jurdjevic1985polynomial} therefore
implies that $(\bar{F},\bar{B})$ is controllable.
Hence, the complete support admits a controllable realization and
\eqref{eq:nonuniform-polynomial-system} is structurally controllable.
The same argument applies to any  driver node set that renders $H_d$
structurally controllable.
\end{proof}

Proposition~\ref{prop:nonuniform-highest-degree} provides a sufficient
but not necessary condition. Lower-degree interactions may introduce
additional controllability directions or accessibility pathways that
are absent from the highest-degree subsystem. Hence, the complete
non-uniform system may be structurally controllable even when $H_d$
is not, and a driver-node set obtained from $H_d$ may be conservative.
Under the standard independent-parameter assumption, this conclusion
is generic: once a controllable realization exists, the corresponding
Lie-algebraic rank condition holds for almost all admissible parameter
values, since the exceptional set is contained in the zero set of a
nontrivial polynomial rank minor.

\subsection{Non-polynomial systems}
For the case of non-polynomial systems, the structural controllability problem becomes much harder as there are no strong controllability results for non-polynomial systems, therefore we take a heuristic approach. 
Analogously to what is observed in \cite{structural_controllability}, we exploit the fact that once all leaf nodes are controlled, there is a directed path from the input signal to any node in the system.  While controlling the leaf nodes ensures  reachability, the no-dilation
condition may not be satisfied. Even when both conditions
hold, they do not guarantee structural controllability for general
non-polynomial systems.

For instance in the left panel of Figure~\ref{fig:loose-ends}, the input signal cal propagate from node 2 to node 1 and from node 3 to node 1. 
However, the input signal cannot propagate to node 2, neither from node 1 nor from node 3. 
All together, both nodes 2 and 3 have to be controlled. 
In the middle panel of Figure~\ref{fig:loose-ends}, the input signal propagate to node 1 and node 3. 
It is therefore enough to control node 2. 
Finally, in the right panel, the input signal propagate from any node to any other. 
Therefore, any of the node can be controlled in order to control the whole system.

Notice that a system could be disconnected, while all nodes are reachable (e.g., two copies of the right panel of Figure~\ref{fig:loose-ends}). 
In such a case, one would need to control at least one node in each of the connected components.

\section{Control Algorithm}\label{se:workflow}
We are now ready to design our system identification and control algorithm. The proposed algorithm follows the following process pipeline, summarized in Figure~\ref{fig:summary}:
\begin{enumerate}
    \item[A.] Initialization and State Space Exploration
    \item[B.] System Identification via THIS
    \item[C.] Droop Controller Design
\end{enumerate}
All steps are explained in detail in the following dedicated subsections.

\subsection{Initialization and State Space Exploration}
The system is initialized in the vicinity of a locally stable equilibrium point ${x}^*$ and recording of the nodes' time series begins at time $t_0=0$.
Given the noise, the system explores the neighborhood of ${x}^*$ and the trajectories, after a certain time interval $[t_0,t_1]$, are rich enough to serve as input to the identification algorithm THIS. 
No control is applied within the time interval $[t_0,t_1]$.
\subsection{System Identification via THIS}
At time $t_1>t_0$, the recorded time series are provided as input to THIS to infer the hypergraph structure. We recall that THIS is a model-free hypergraph inference algorithm designed to uncover causal relationships among system variables. It enables robust hypergraph reconstruction with minimal prior knowledge about the underlying dynamics, requiring only that the system can be represented as a set of coupled differential equations.

In terms of the exploration domain, noise plays a constructive role in THIS: the explored region should be sufficiently large to capture the system’s nonlinear behavior (beyond the locally linear regime), yet not excessively broad to maintain the validity of the Taylor approximation, which is crucial for THIS. When this balance is achieved, THIS remains fully agnostic to the specific regions of the state space from which data are collected.

The run time of THIS depends on the size of the system and on the length of the time series considered. 
In our pipeline, THIS runs from $t_1$ to $t_2$, while the system evolves freely.

\subsection{Droop Controller Design}
 Given the inferred hypergraph, a  droop control is designed and applied to the nodes in the minimal set $S^*$ of controllable nodes. 
The state-feedback control  follows a linear proportional control law $u_k = K(x_k-x_k^*)$ with $k\in S^* $ and $K$ chosen such that stability of ${x}^*$ is preserved.
At time instant $t_2\geq t_1$, the droop control is turned on and noise is damped throughout the system.
The effectiveness of this control scheme is showcased in Section~\ref{sec:numerics}.

\section{Numerical Simulations}\label{sec:numerics}
We apply our inference-control algorithm to both a polynomial system (the generalized Lotka-Volterra model of population dynamics~\cite{goel1971volterra}) and to a non-polynomial system (the higher-order Kuramoto model). 

\subsection{The generalized Lotka-Volterra model}
The abundances' evolution of $n$ species in an ecosystem is typically modeled by the \emph{generalized Lotka-Volterra model}~\cite{goel1971volterra}:
\begin{align}
\begin{split}
    \dot{x}_i &= r_ix_i\left(1-\frac{x_i}{l_i}\right) + \sum_{j_1=1}^{n}{A}_{ij_1}^{(2)}x_ix_{j_1} \\
    & + \sum_{1\leq j_1<j_2\leq n}{A}_{ij_1j_2}^{(3)}x_ix_{j_1}x_{j_2} + \cdots
\end{split}
\end{align}
where $r_i$ and $l_i$ are the intrinsic groth rate and carrying capacity of species $i$, and ${A}^{(k)}$ is the adjacency tensor of $k$-th order interactions. 
We indicate with ${x}^*$ the steady state.

\subsection{Higher-order Kuramoto model}
Let us consider the generalization of the Kuramoto model to higher-order interactions, as a special case of \eqref{eq:system}. The dynamics of the $i$-th phase is governed by
\begin{equation}
\begin{split}
    \dot{x}_i &= \omega_i + \sum_{j=1}^{n} A^{(2)}_{ij} \sin(x_j-x_i) \\
    &+ \sum_{j,k=1}^{n} A^{(3)}_{ijk} \sin(x_j+x_k-2x_i) + \cdots
\label{eq:kuramoto}
\end{split}
\end{equation}
where $x_i \in S^1$ represents the phase of oscillator $i$ and $\omega_i\in \mathbb{R}$ is its natural frequency. 
We indicate with ${x}^*$ its synchronous state. 

\begin{figure*}[!ht]
    \centering
    \includegraphics[width=\textwidth]{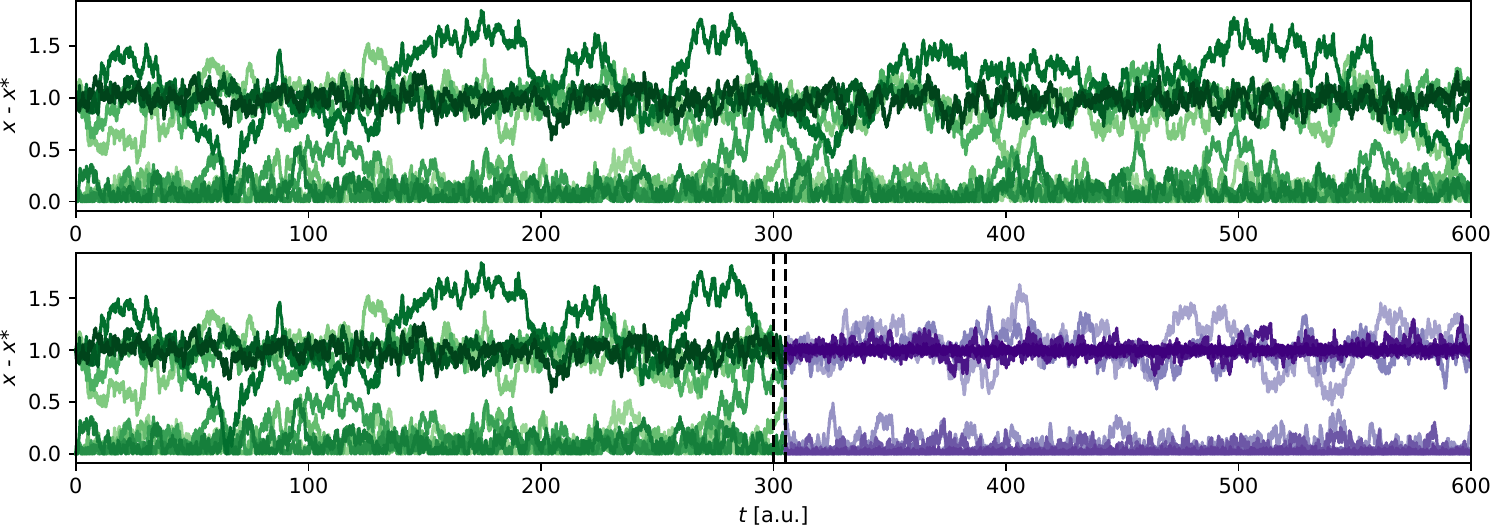}
    \caption{Top panel: Trajectory of the 10 agents of a 3rd-order generalized Lotka-Volterra model on a random 3-hypergraph. 
    Each agent deviates from its steady state due to noise. 
    Bottom panel: The time series of the nodes over the first 300 time units are used to infer the hypergraph structure using THIS. 
    Within this hypergraph, the driver nodes are identified and a proportional control is applied to them, starting at time $t_2=305$.
    From $t_0=0$ to $t_2=305$, the two time series are exactly the same.}
    \label{fig:droop-lv}
\end{figure*}

\begin{figure*}[!ht]
    \centering
    \includegraphics[width=\textwidth]{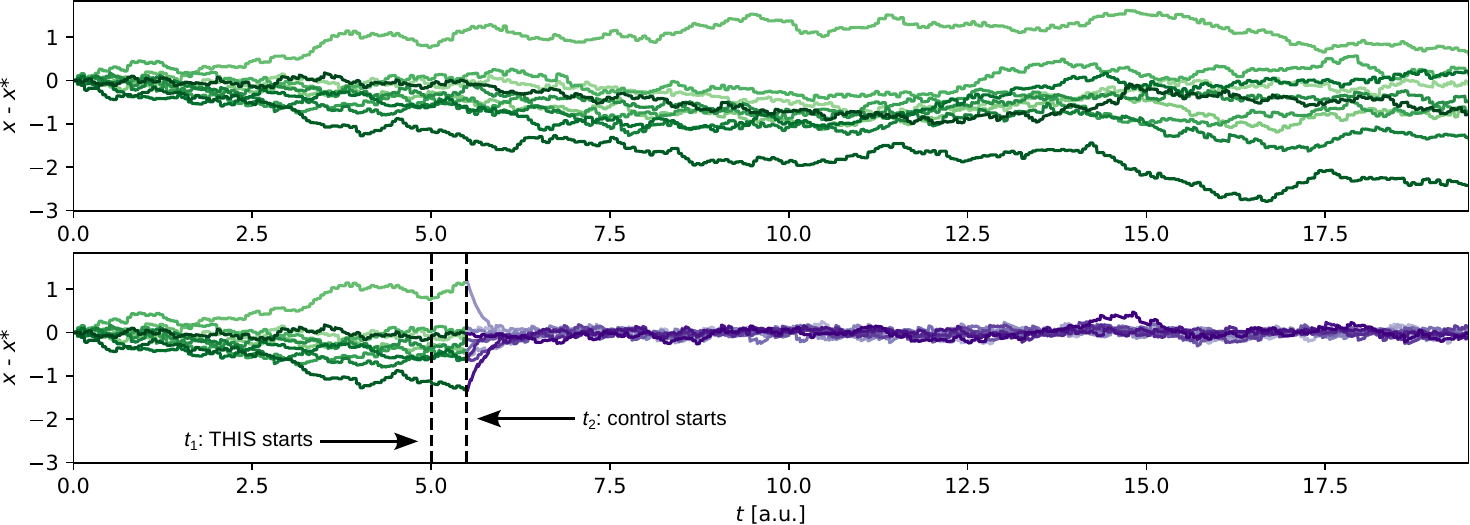}
    \caption{Top panel: Trajectory of the 10 agents of a 3rd-order Kuramoto model on a random 3-hypergraph. 
    Each agent deviates from its steady state due to noise. 
    Bottom panel: The time series of the nodes over the first 5 time units are used to infer the hypergraph structure using THIS. 
    Within this hypergraph, the driver nodes are identified and a proportional control is applied to them, starting at time $t_2=5.5$.
    From $t_0=0$ to $t_2=5.5$, the two time series are exactly the same.}
    \label{fig:this-control}
\end{figure*}

\subsection{Simulation results}
We illustrate our results for both systems with third-order hypergraph of interactions among $n$ dynamical nodes, i.e., $\mathcal{H}(V,E,A^{(2)},A^{(3)})$. 

In our example, the systems are made up of $n=10$ nodes interacting in a random third order hypergraph. 
As shown in Figures~\ref{fig:droop-lv} and \ref{fig:this-control} each state deviates from the steady state due to noise. 
After the state space exploration in the time interval $t\in[0,t_1]$ all the state space trajectories are collected and the identification step is initiated. 
THIS infers the hypergraphs in the time interval $t~\in~[t_1,t_2]$. 
Here we virtually allocate an arbitrary time to the inference for illustrative purpose. 
In practice, the inference time strongly depends on the system size and the time series length. 
Starting at $t_2$, the proportional control law $u_k=K(x_k-x_k^*)$ with $K=5$ is applied to each driver node identified by THIS, which in turn drive the overall system towards the desired state.

Interestingly, we notice that THIS does not infer exactly the actual hypergraph underlying the dynamics. 
Indeed, in this particular example, THIS achieves a true positive rate of between 25\% and 55\% on the inference of 3-edges. 
At the same time, THIS infers some 2-edges, while there was none in the actual hypergraph. 
Nevertheless, THIS appropriately identifies a set of driver nodes and hence does a good job in designing an effective controller. 
More quantitatively, the $H_2$-norm of the deviation from steady state is reduced by at least one order of magnitude compared to the uncontrolled system. 

This imperfect inference is generic to our examples and is not an artifact of a particular simulation. 
The main reason for these errors in the inferred edges stems from the limited radius of exploration allowed by the chosen noise, over the chosen time interval. 
The inference would have been better should the magnitude of the noise be larger, or the exploration time be longer.

\section{Conclusions}\label{sec:conclusions}
We proposed an explicit implementation of a data-driven controller on hypernetwork systems. 
Based on measured time series of the system's nodes, THIS infers the underlying hypernetwork structure and designs a minimal controller accordingly. 

So far, the proposed approach performs best on diffusively coupled systems where a stable steady state exists (e.g., the Kuramoto model). 
The strength of the approach lies in the fact that it is mostly system agnostic and does not need any prior on the studied system. 

Most interestingly, it appears that, despite a partially incorrect inference of the hypergraph, the set of effective control nodes is perfectly identified.
This last observation opens promising avenues for the application of (hyper)network inference. 
Indeed, it appears that inference algorithms like THIS are capable of extracting the key information needed for the hypergraph controllability, despite model mismatches.


\end{document}